\documentclass{aa}

\usepackage{amsmath}
\usepackage{bm}
\usepackage{commath}
\usepackage{gensymb}
\usepackage{graphicx}
\usepackage{natbib}
\usepackage{placeins}
\usepackage{subcaption}
\usepackage[flushleft]{threeparttable}
\usepackage{txfonts}
\usepackage{textcomp}
\usepackage{upgreek}
\bibpunct{(}{)}{;}{a}{}{,}

\begin{document}

   \title{A dynamical and radiation semi-analytical model of pulsar-star colliding winds along the orbit: Application to LS~5039}
   \titlerunning{A model for pulsar-star colliding winds}

   \author{E. Molina \and V. Bosch-Ramon}

   \institute{Departament de F\'isica Qu\`antica i Astrof\'isica,
              Institut de Ci\`encies del Cosmos (ICCUB),
              Universitat de Barcelona (IEEC-UB), 
              Mart\'i i Franqu\`es 1, 08028 Barcelona, Spain \\
              \email{emolina@fqa.ub.edu, vbosch@fqa.ub.edu}}

   \date{Received -; accepted -}

  \abstract 
   {Gamma-ray binaries are systems that emit non-thermal radiation peaking at energies above 1~MeV. One proposed scenario to explain their emission consists of a pulsar orbiting a massive star, with particle acceleration taking place in shocks produced by the interaction of the stellar and pulsar winds.}
   {We develop a semi-analytical model of the non-thermal emission of the colliding-wind structure including the dynamical effects of orbital motion. We apply the model to a general case and to LS~5039.}
   {The model consists of a one-dimensional emitter the geometry of which is affected by Coriolis forces owing to orbital motion. Two particle accelerators are considered: one at the two-wind standoff location, and the other one at the turnover produced by the Coriolis force. Synchrotron and inverse Compton emission is studied, accounting for Doppler boosting and absorption processes associated to the massive star.}
   {If both accelerators are provided with the same energy budget, most of the radiation comes from the region of the Coriolis turnover and beyond, up to a few orbital separations from the binary system. Significant orbital changes of the non-thermal emission are predicted in all energy bands. The model allows us to reproduce some of the LS~5039 emission features, but not all of them. In particular, the MeV radiation is probably too high to be explained by our model alone, the GeV flux is recovered but not its modulation, and the radio emission beyond the Coriolis turnover is too low. The predicted system inclination is consistent with the presence of a pulsar in the binary.}
   {The model is quite successful in reproducing the overall non-thermal behavior of LS~5039. Some improvements are suggested to better explain the phenomenology observed in this source, like accounting for particle reacceleration beyond the Coriolis turnover, unshocked pulsar wind emission, and the three-dimensional extension of the emitter.}

   \keywords{gamma-rays: stars - radiation mechanisms: non-thermal - stars: winds, outflows - stars: individual: LS~5039}
   \maketitle

\section{Introduction}\label{introduction}

Gamma-ray binaries are binary systems consisting of a compact object, which can be a black hole or a neutron star, and a non-degenerate star. These sources emit non-thermal radiation from radio up to very high-energy gamma rays (VHE; above 100~GeV), and the most powerful ones host massive stars \citep[see][for a review]{dubus13,paredes19}. The main difference with X-ray binaries, which may also show persistent or flaring gamma-ray activity \citep[see, e.g.,][for Cygnus~X-1 and Cygnus~X-3, respectively, and references therein]{zanin16,zdziarski18}, is that in X-ray binaries emission reaches its maximum at X-rays, whereas gamma-ray binaries emit most of their radiation at energies above 1~MeV. 

Up to date, there are nine confirmed high-mass gamma-ray binaries for which emission above 100~MeV has been detected: LS~I~+61~303 \citep{tavani98}, LS~5039 \citep{paredes00}, PSR~B1259-63 \citep{aharonian05}, HESS~J0632+057 \citep{hess07,hinton09}, 1FGL~J1018.6-5856 \citep{fermi12}, HESS~J1832-093 \citep{hess15,eger16}, LMC~P3 \citep{corbet16}, PSR~J2032+4127 \citep{veritas18}, and 4FGL~J1405-6119 \citep{corbet19}. There are other candidate systems with a pulsar orbiting a massive star that exhibit non-thermal radio emission, but for which gamma rays have yet to be detected \citep[see][and references therein]{dubus17}.

The two most common scenarios proposed to explain the observed non-thermal emission in gamma-ray binaries involve either a microquasar, which generates non-thermal particles in relativistic jets powered by accretion onto a compact object \citep[see, e.g.,][for a thorough study of the scenario]{bosch09a}, or a non-accreting pulsar, in which energetic particles are accelerated through shocks produced by the interaction of the stellar and pulsar winds \citep[e.g.][]{maraschi81,leahy04,dubus06,khangulyan07,sierpowska07,zabalza13,takata14,dubus15}. With the contribution of orbital motion at large scales, this wind interaction leads to the formation of a spiral-like structure composed mainly of shocked pulsar material that can extend up to several dozen times the orbital separation \citep[see, e.g.,][for an analytical description and numerical simulations of this structure]{bosch11,bosch12,bosch15,barkov16}.

In this work, we present a model to describe the broadband non-thermal emission observed in gamma-ray binaries through the interaction of the stellar and pulsar winds. The novelty of this model with respect to previous works is the application of a semi-analytical hydrodynamical approach to study the combined effect of the stellar wind and orbital motion on the emitter, which is assumed to be one-dimensional (1D) \citep[see][for a similar model in a microquasar scenario]{molina18,molina19}. The paper is structured as follows: the details of the model are given in Sect.~\ref{model}, the results for a generic system are shown in Sect.~\ref{results}, a specific application to the case of LS~5039 is done in Sect.~\ref{LS5039}, and a summary and a discussion are given in Sect.~\ref{discussion}.

\section{Description of the general model}\label{model}

\begin{figure}
    \centering
    \includegraphics[width=\linewidth]{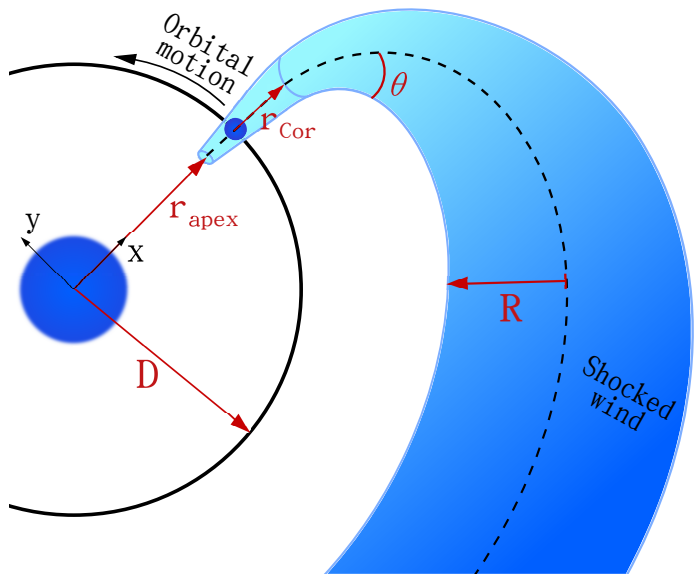}
    \caption{Schematic zenithal view of the studied scenario (not to scale). Only a fraction of the orbit is shown for clarity.}
    \label{fig:sketch}
\end{figure}

As a representative situation, we study a binary system made of a massive O-type star and a pulsar that orbit around each other with a period of $T = 5$~days. The orbit is taken circular for simplicity, with an orbital separation of $D = 3\times 10^{12}$~cm $\approx 0.2$~AU. The stellar properties are typical for a main sequence O-type star \citep{muijres12}, namely a temperature of $T_\star = 40,000$~K and a luminosity of $L_\star = 10^{39}$~erg~s$^{-1}$. For simplicity, we model the stellar wind as an isotropic supersonic outflow with a velocity of $v_{\rm w} = 3\times10^8$~cm~s$^{-1}$ and a mass-loss rate of $\dot{M}_{\rm w} = 3\times10^{-7}$~M$_\sun$~yr$^{-1}$. The wind velocity is taken constant and not following the classical $\beta$-law for massive stars \citep[e.g.][]{pauldrach86}, as the effect of considering such a velocity profile for standard $\beta$ values is small for the purpose of this work. The pulsar has a spin-down luminosity of $L_{\rm p} = 3\times10^{36}$~erg~s$^{-1}$, which is taken here as equal to the kinetic luminosity of the pulsar wind. The latter is assumed to be ultra-relativistic and isotropic, with Lorentz factor $\Gamma_{\rm p} = 10^5$. The distance to the system is taken to be $d = 3$~kpc, and the inclination $i$ is left as a free parameter. A list of the parameter values used for a generic gamma-ray binary can be found in Table~\ref{tab:parameters}, and a sketch of the studied scenario is presented in Fig.~\ref{fig:sketch}. Throughout this section, the notation $u = ||\overrightarrow{u}||$ is used to refer to a vector norm. Also, we will use primed quantities in the fluid frame (FF), and unprimed ones in the laboratory frame (LF) of the star.

\begin{table}
 \begin{center}
	\caption{List of the parameters that are used in this work. The last three are those for which different values are explored.}
	\begin{tabular}{c l c}
    \hline \hline
	\multicolumn{1}{c}{}    & \multicolumn{1}{c}{Parameter} & Value                                 \\
    \hline
    \rule{0pt}{2.3ex}
	Star        & Temperature $T_\star$		   	            & $4\times10^4$~K                       \\
                & Luminosity $L_\star$ 		                & $10^{39}$~erg~s$^{-1}$		        \\
                & Mass-loss rate $\dot{M}_{\rm w}$          & $3\times10^{-7}$~M$_\sun$~yr$^{-1}$	\\
                & Wind speed $v_{\rm w}$  	        	    & $3\times10^8$~cm~s$^{-1}$		        \\
    \hline
    \rule{0pt}{2.3ex}
    Pulsar      & Luminosity $L_{\rm p}$ 	            	& $3\times10^{36}$~erg~s$^{-1}$         \\
                & Wind Lorentz factor $\Gamma_{\rm p}$      & $10^5$                                \\
    \hline
    \rule{0pt}{2.3ex}
    System      & Orbital separation $D$ 		            & $3\times10^{12}$~cm                   \\
                & Orbital period $T$ 				        & $5$~days                              \\
                & Orbital eccentricity $e$                  & $0$                                   \\
                & Distance to the observer $d$ 				& $3$~kpc                               \\
                & Non-thermal fraction $\eta_{\rm NT}$ 	    & $0.1$                                 \\
                & Acceleration efficiency $\eta_{\rm acc}$  & $0.1$                                 \\
                & Injection power-law index $p$             & $-2$                                   \\
                & Coriolis turnover speed $v_{\rm Cor}$     & $3\times10^9$, $10^{10}$~cm~s$^{-1}$  \\
                & Magnetic fraction $\eta_B$                & $10^{-3}$, $10^{-1}$                  \\
                & System inclination $i$                    & $30\degree$, $60\degree$              \\
    \hline
    \end{tabular}
    \label{tab:parameters}
 \end{center}
\end{table}

\subsection{Dynamics}\label{dynamics}

The interaction of the stellar and the pulsar winds produces an interface region between the two where the shocked flow pressures are in equilibrium: the so-called contact discontinuity (CD). Close to the binary system, where the orbital motion can be neglected, the shape of this surface is characterized by the pulsar-to-stellar wind momentum rate ratio, defined as
    \begin{equation}\label{eta}
    \eta = \frac{L_{\rm p}}{\dot{M}_{\rm w} v_{\rm w} c}
    \end{equation}
for $\Gamma_{\rm p} \gg 1$. The asymptotic half-opening angle of the CD can be approximated by the following expression from \cite{eichler93}, which is in good agreement with numerical simulations \citep[e.g.][]{bogovalov08}:
    \begin{equation}\label{theta}
    \theta = 28.6\degree (4 - \eta^{2/5}) \eta^{1/3} \ .
    \end{equation}
The apex of the CD, where the two winds collide frontally, is located along the star-pulsar direction at a distance from the star of $r_{\rm apex} = D/(1 + \sqrt{\eta})$, with $D$ being the orbital separation. For the adopted parameters, we obtain $\eta = 0.018$, $\theta = 28.7\degree$, and $r_{\rm apex} = 0.88D = 2.6\times10^{12}$~cm, the latter being constant owing to the circularity of the orbit. 

The star is located at the origin of the coordinate system, which co-rotates with the pulsar. The $x$-axis is defined as the star-pulsar direction, and the $y$-axis is perpendicular to it and points in the direction of the orbital motion. We model the evolution of the shocked stellar and pulsar winds in the inner interaction region as a straight conical structure of half-opening angle $\theta$ and increasing radius $R$, the onset of which is at $x = r_{\rm apex}$, where it has a radius of $R_0 = D - r_{\rm apex}$, roughly corresponding to the characteristic size of the CD at its apex. The shocked flows are assumed to move along the $x$ direction in this region (i.e., the orbital velocity is neglected here with respect to the wind speed), until they reach a point where their dynamics start to be dominated by orbital effects, the Coriolis turnover. Beyond this point, the CD is progressively bent in the $-y$ direction due to the asymmetric interaction with the stellar wind, which arises from Coriolis forces (see Fig.~\ref{fig:sketch} for an schematic view). As a result, the shocked flow structure acquires a spiral shape at large scales. One can estimate the distance of the Coriolis turnover to the pulsar $r_{\rm Cor}$ by following the analytical prescription in \cite{bosch11}, which comes from equating the total pulsar wind pressure to the stellar wind ram pressure due to the Coriolis effect (for $\Gamma_{\rm p} \gg 1$):
     \begin{equation}\label{rCor}
     \frac{L_{\rm p}}{4\pi c r_{\rm Cor}^2} = \frac{\rho_{\rm w}(D)}{(1 + r_{\rm Cor}/D)^2} \left( \frac{4 \pi}{T} \right)^2 r_{\rm Cor}^2 \ ,
     \end{equation}
where $\rho_{\rm w}(r) = \dot{M}_{\rm w}/4\pi r^2 v_{\rm w}$ is the stellar wind density at a distance r from the star. Although approximate, this expression agrees well with results from 3D numerical simulations \citep{bosch15}. Numerically solving Eq.~\eqref{rCor} for our set of parameters yields $r_{\rm Cor} = 0.94D$. Note that in the case of an elliptical system $r_{\rm Cor}$ changes along the orbit.

The radiation model explained in Sect.~\ref{emitter} focuses on the shocked pulsar material flowing inside the CD. The values of the Lorentz factor of this fluid are set to increase linearly with distance from the CD apex ($\Gamma_0 = 1.06, v_0 = c/3$) to the Coriolis turnover ($\Gamma = 4, v = 0.97c$), based on the simulations by \cite{bogovalov08}. Beyond the Coriolis turnover, and due to some degree of mixing between the pulsar and stellar winds \citep[see, e.g., the numerical simulations in][]{bosch15}, we fix the shocked pulsar wind speed $v_{\rm Cor}$ to a constant value that is left as a free parameter in our model, and could account for different levels of mixing. The shocked stellar wind that surrounds the shocked pulsar wind plays the role of a channel through which the former flows, and may have a much lower speed. This channel is assumed to effectively transfer the lateral momentum coming from the unshocked stellar wind to the shocked pulsar wind.

The trajectory of the shocked winds flowing away from the binary, which are assumed to form a (bent) conical structure, is determined by accounting for orbital motion and momentum balance between the upstream shocked pulsar wind and the unshocked stellar wind, with a thin shocked stellar wind channel playing the role of a mediator. This conical structure is divided into 2000 cylindrical segments of length $dl = 0.05D$, amounting to a total length of $100D$. Initially, all the shocked material moves along the $x$ axis between $x = r_{\rm apex}$ and $x = D + r_{\rm Cor}$. At the latter point, its position and momentum are, in Cartesian coordinates of the co-rotating frame,
     \begin{equation}\label{initialC}
     \begin{aligned}
     \overrightarrow{r} &= (D + r_{\rm Cor},0) \\
     \overrightarrow{P} &= (\dot{P}{\rm d}t,0) \ ,
     \end{aligned}
     \end{equation}
where $\dot{P}$ is the momentum rate of the shocked flow, which we estimate as:
     \begin{equation}\label{Pdot}
     \dot{P} = \frac{L_{\rm p}}{c} \ .
     \end{equation}
For simplicity, we are not considering the contribution of the momentum of the stellar wind loaded through mixing into the pulsar wind before the Coriolis turnover. This contribution depends on the level of mixing, and could be as high as $\Omega \dot{M}_{\rm w} v_{\rm w}$, with $\Omega = R^2 / 4 (D + r_{\rm Cor})^2$ being the solid angle fraction subtended by the shocked pulsar wind at the Coriolis turnover, as seen from the position of the star. We note, nonetheless, that the inclusion of the stellar wind contribution to $\dot{P}$ has a modest impact on the radiation predictions of the model and can be neglected at this stage.

The unshocked stellar wind velocity in the co-rotating frame has components in both the $x$ and the $y$ directions:
     \begin{equation}\label{vw}
     \overrightarrow{\hat{v}_{\rm w}} = v_{\rm w} \frac{(x,y)}{r} + \omega r \frac{(y,-x)}{r} \ ,
     \end{equation}
where $\omega$ is the orbital angular velocity, and the hat symbol is used to distinguish $\hat{v}_{\rm w}$ from the purely radial component of the wind, $v_{\rm w}$. In a stationary configuration, the force that the unshocked stellar wind exerts onto each segment of the shocked wind structure is
     \begin{equation}\label{Fw}
     \overrightarrow{F_{\rm w}} = \rho_{\rm w} \hat{v}_{\rm w}^2 S \sin{\alpha} \frac{\overrightarrow{\hat{v}_{\rm w}}}{\hat{v}_{\rm w}} \ ,
     \end{equation}
with $\alpha$ being the angle between $\overrightarrow{\hat{v}_{\rm w}}$ and the shocked pulsar wind velocity $\overrightarrow{v}$, and $S\sin\alpha = 2 R dl\sin\alpha$ the segment lateral surface normal to the stellar wind direction. The component of $\overrightarrow{F_{\rm w}}$ parallel to the fluid direction is assumed to be mostly converted into thermal pressure, whereas the perpendicular component $\overrightarrow{F_{\rm w}^\perp} = \overrightarrow{F_{\rm w}} \sin{\alpha}$ modifies the segment momentum direction. Thus, the interaction with the stellar wind only reorients the fluid, but it does not change its speed beyond the Coriolis turnover\footnote{There must be some acceleration of the shocked flow away from the binary as a pressure gradient is expected, but for simplicity this effect is neglected here.}. 

We find the conditions for the subsequent segments by applying the following recursive relations:
     \begin{equation}\label{iterative}
     \begin{aligned}
     \overrightarrow{P_{\rm i}}_{+1} &= \overrightarrow{P_{\rm i}} + \overrightarrow{F_{\rm w}^\perp} dt_{\rm i} \\
     \overrightarrow{v_{\rm i}}_{+1} &= v_{\rm i} \frac{\overrightarrow{P_{\rm i}}}{P_{\rm i}} \\
     \overrightarrow{r_{\rm i}}_{+1} &= \overrightarrow{r_{\rm i}} + \overrightarrow{v_{\rm i}} dt_{\rm i} \ ,
     \end{aligned}
     \end{equation}
where $v_{\rm i}$ is the shocked pulsar wind velocity in each segment, and $dt_{\rm i} = dl/v_{\rm i}$ is the segment advection time. This procedure yields a fluid trajectory semi-quantitatively similar to that obtained from numerical simulations within approximately the first spiral turn \citep[e.g.][]{bosch15}.

\subsection{Characterization of the emitter}\label{emitter}

The non-thermal emission is assumed to take place in the shocked pulsar wind, which moves through the shocked stellar wind channel and follows the trajectory defined in the previous section. We only consider a non-thermal particle population consisting only of electrons and positrons. These are radiatively more efficient than accelerated protons \citep[e.g.][]{bosch09a}, but the presence of the latter cannot be discarded. Particles are accelerated at two different regions where strong shocks develop: the pulsar wind termination shock, located here at the CD apex; and the shock that forms in the Coriolis turnover \citep[as in][hereafter the Coriolis shock]{zabalza13}. In our general model, we consider for simplicity that both regions have the same power injected into non-thermal particles (in the LF), taken as a fraction of the total pulsar wind luminosity: $L_{\rm NT} = \eta_{\rm NT} L_{\rm p}$, with $\eta_{\rm NT} = 0.1$. They also have the same acceleration efficiency $\eta_{\rm acc} = 0.1$, which defines the energy gain rate $\dot{E}'_{\rm acc} = \eta_{\rm acc} e c B'$ of particles for a given magnetic field $B'$. The latter is defined through its energy density being a fraction of the total energy density at the CD apex (indicated with the subscript 0):
    \begin{equation}\label{etaB}
    \frac{B_0^{\prime 2}}{8\pi} = \eta_B \frac{L_{\rm p}}{\pi R_0^2 v_0 \Gamma_0^2} \ .
    \end{equation}
The magnetic field is assumed toroidal, and therefore it evolves along the emitter as $B' \propto R^{-1} \Gamma^{-1}$. For simplicity, we assume that all the flow  particles at the Coriolis turnover are reprocessed by the shock there, leading to a whole new population of particles. This means that the particle population beyond the Coriolis shock only depends on the properties of the latter, and is independent of the particle energy distribution coming from the initial shock at the CD apex. This assumption divides the emitter into two independent and distinct regions: one region between the CD apex and the Coriolis turnover (hereafter the inner region), and one beyond the latter (hereafter the outer region). We recall that the inner region has a velocity profile corresponding to a linear increase in $\Gamma$, whereas in the outer region the fluid moves at a constant speed (see Sect.~\ref{dynamics}).

To compute the particle evolution, the emitter is divided into 1000 segments of length $0.1D$, in order to account for the same total length as in Sect~\ref{dynamics}. An electron and positron population is injected at each accelerator following a power-law distribution in the energy $E'$, with an exponential cutoff and spectral index $p$:
    \begin{equation}\label{Q}
    Q'(E') \propto E'^p \exp{\left( -\frac{E'}{E'_{\rm max}}\right)} \ ,
    \end{equation}
where $E'_{\rm max}$ is the cutoff energy obtained by comparing the acceleration timescale, $t'_{\rm acc} = E'/|\dot{E}'_{\rm acc}|$, with the cooling and diffusion timescales, $t'_{\rm cool} = E'/|\dot{E}'|$ and $t'_{\rm diff} = 3 R'^2 e B' / 2 c E'$, respectively ($R$ is the perpendicular size and thus is taken $R'=R$). We adopt $p = -2$ because it allows for a substantial power to be available for gamma-ray emission. Harder, and also a bit softer electron distributions would also be reasonable options. The particle injection is normalized by the total available power $L'_{\rm NT} = L_{\rm NT}/\Gamma^2$. We note that $E'_{\rm max}$ and $L'_{\rm NT}$ are not the same for both accelerators, since their properties differ. Particles are advected between subsequent segments following the bulk motion of the fluid \citep[see Appendix B2 in][]{delacita16}, and they cool down via adiabatic, synchrotron, and inverse Compton (IC) losses as they move along the emitter. The particle energy distribution at each segment is computed in the FF following the same recursive method as in \cite{molina18}, which yields, for a given segment k:
    \begin{equation}\label{NE}
    N'_{\rm k}(E'_{\rm k}) = N'_0(E'_0) \prod_{\rm i=k}^1 \frac{\dot{E}'_{\rm i}(E'_{{\rm i}-1})}{\dot{E}'_{\rm i}(E'_{\rm i})} \ ,
    \end{equation}
where $E'_{\rm k}$ is the energy of a given particle at the location of segment $k$, and $E'_{\rm i}$ is the energy that this same particle had when it was at the position of segment $i$, with $i \le k$ (we note that $E'_{\rm i} > E'_{\rm k}$ due to energy losses).

For every point where we have the particle distribution, the synchrotron spectral energy distribution (SED) is computed following \cite{pacholczyk70} for an isotropic distribution of electrons in the FF. The IC SED is obtained from the numerical prescription developed by \cite{khangulyan14} for a monodirectional field of stellar target photons with a black-body spectrum. These SEDs are then corrected by Doppler boosting and absorption processes, the latter consisting of gamma-gamma absorption with the stellar photons \citep[e.g][]{gould67}, and free-free absorption with the stellar wind ions \citep[e.g.][]{rybicki86}. We do not consider emission from secondary particles generated via the interaction of gamma rays with stellar photons, although it may have a non-negligible impact on our results (see discussion in Sect.~\ref{LS5039Discussion}). Partial occultation of the emitter by the star is also taken into account, although it is only noticeable for very specific system-observer configurations. For a more detailed description of the SED computation we refer the reader to \cite{molina19}.

\section{General results}\label{results}

For the results presented in this section, we make use of the parameter values listed in Table~\ref{tab:parameters}. The orbital phase $\Phi$ is defined such that the pulsar is in the inferior conjunction (INFC) for $\Phi = 0$, and in the superior conjunction (SUPC) for $\Phi = 0.5$.

\subsection{Energy losses and particle distribution}\label{edist}

\begin{figure}
    \centering
    \includegraphics[angle=270,width=\linewidth]{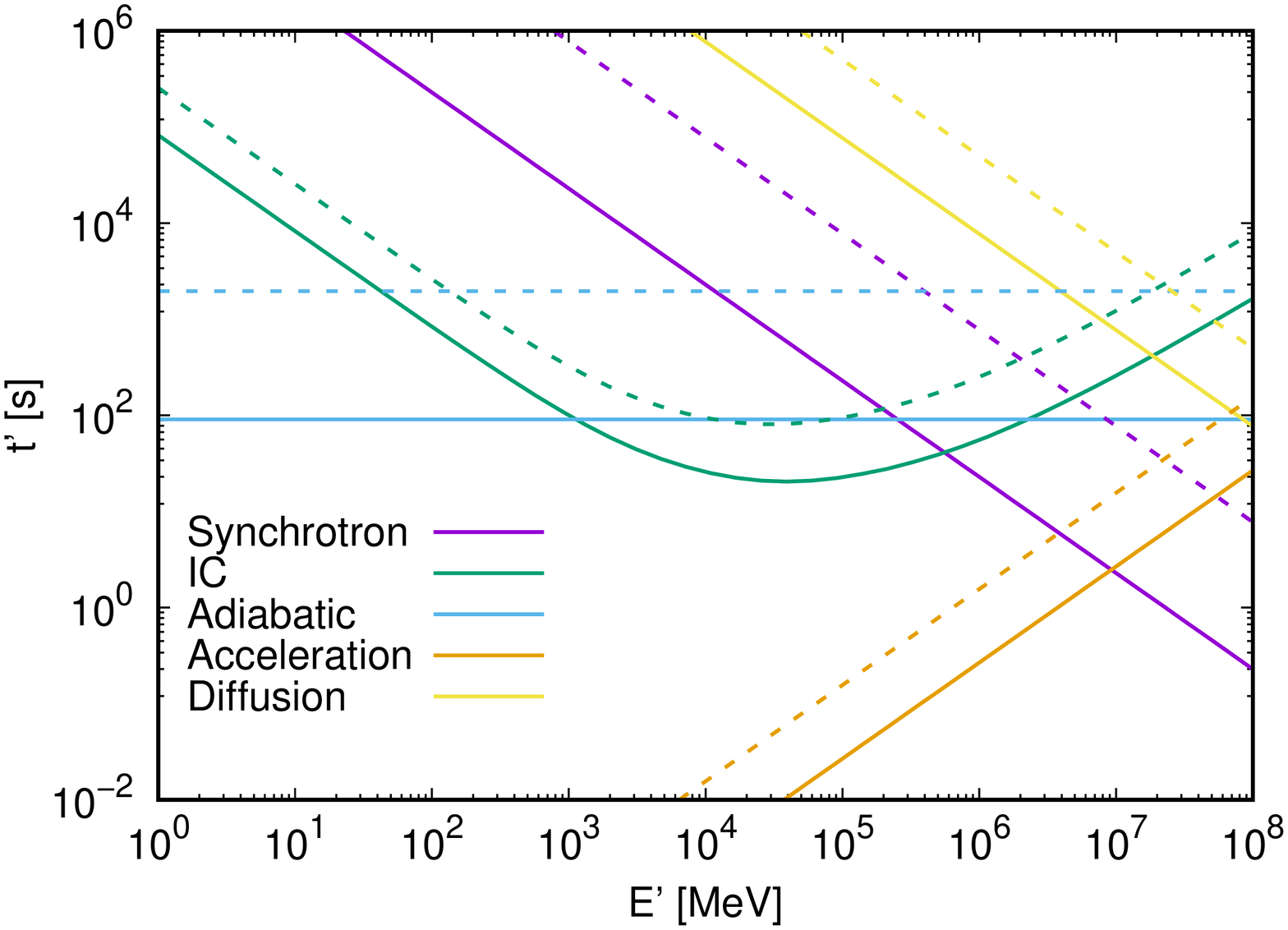}\\
    \vspace{3mm}
    \includegraphics[angle=270,width=\linewidth]{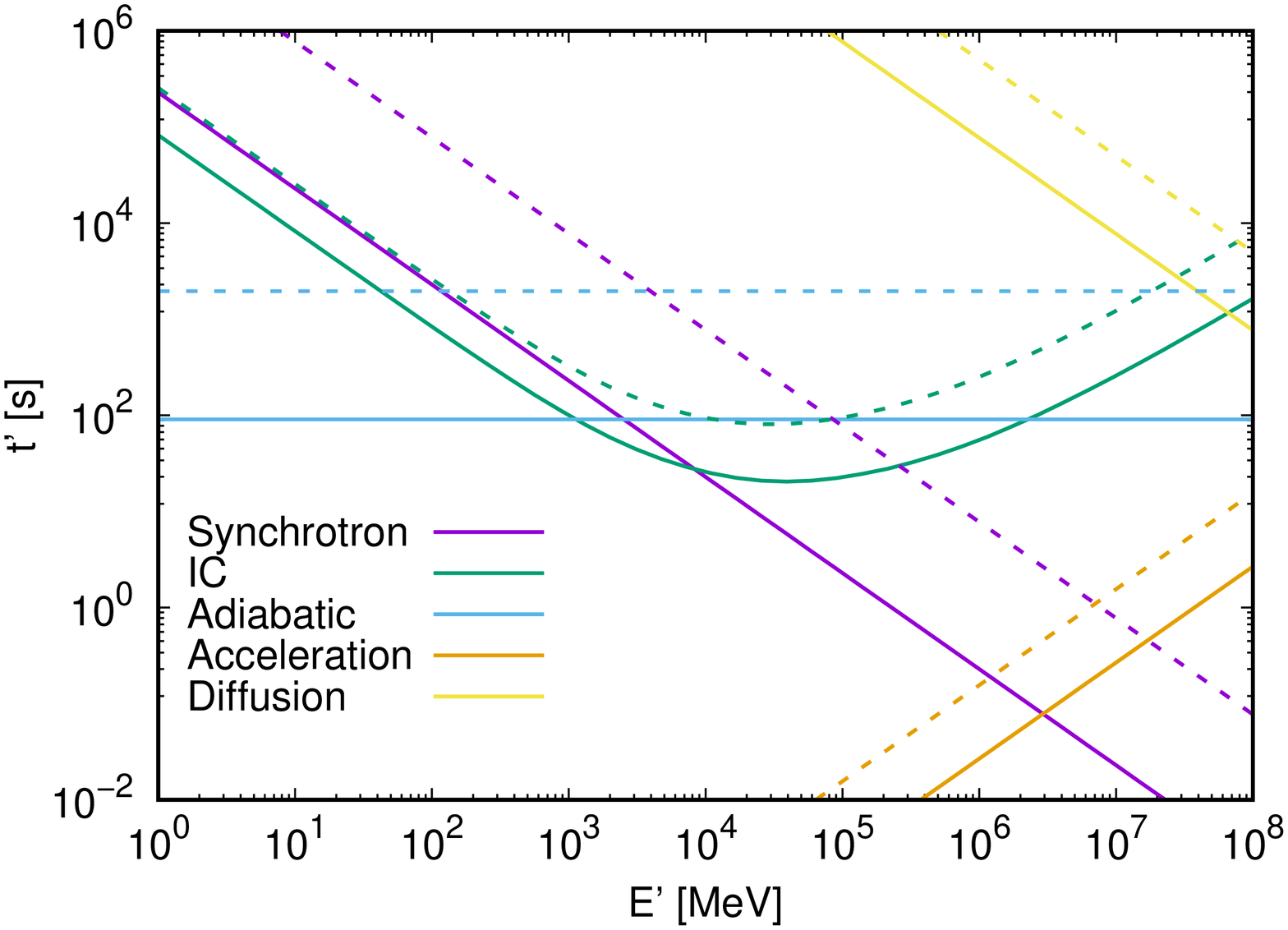}
    \caption{Characteristic timescales in the FF for $v_{\rm Cor} = 3\times10^9$, and $\eta_B = 10^{-3}$ (\textit{top panel}) and $10^{-1}$ (\textit{bottom panel}). Solid and dashed lines represent the values at the CD apex and the Coriolis turnover locations, respectively.}
    \label{fig:losses}
\end{figure}

\begin{figure}
    \centering
    \includegraphics[angle=270,width=\linewidth]{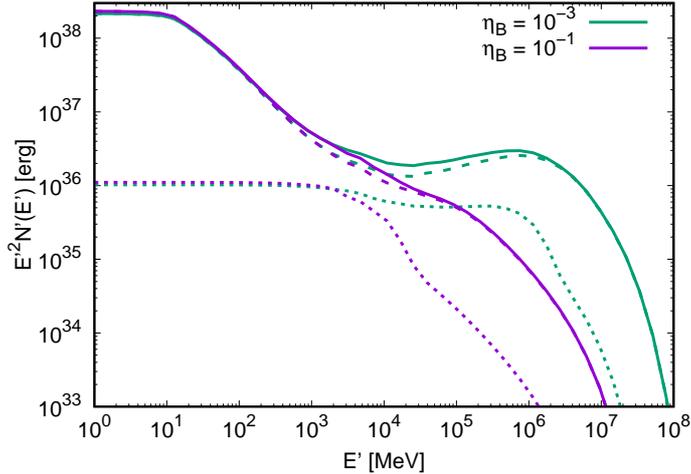}
    \caption{Particle energy distribution in the FF for $v_{\rm Cor} = 3\times10^9$~cm~s$^{-1}$, and $\eta_B = 10^{-3}$ (green lines) and $10^{-1}$ (purple lines). The contributions of the inner and outer regions are represented, respectively, with dotted and dashed lines, and their sum is shown by the solid lines.}
    \label{fig:NE}
\end{figure}

Figure~\ref{fig:losses} shows the characteristic timescales in the FF for the cooling, acceleration, and diffusion processes, for $v_{\rm Cor} = 3\times10^9$~cm~s$^{-1}$, and $\eta_B = 10^{-3}$ and $10^{-1}$, which correspond to initial magnetic fields of $B'_0 = 4.15$~ G and $41.5$~G, respectively. In general, particle cooling is dominated by adiabatic losses at the lowest energies, IC losses at intermediate energies, and synchrotron losses at the highest ones unless a very small magnetic field with $\eta_B < 10^{-5}$ is assumed. The exact energy values at which the different cooling processes dominate depend on $\eta_B$, and also on whichever emitter region we are looking at. Given the dependency of the synchrotron and acceleration timescales on the magnetic field, and that the latter decreases linearly with distance, $E'_{\rm max}$ is higher at the Coriolis turnover than at the CD apex (see where the synchrotron and acceleration lines intersect in Fig.~\ref{fig:losses}), allowing particles to reach higher energies at the location of the former. The larger region size involved and longer accumulation time, combined with the lower energy losses farther from the star, make the non-thermal particle distribution to be dominated by the outer region of the emitter, with just a small contribution from the inner part at middle energies, as seen in Fig.~\ref{fig:NE} (we recall that both regions are assumed to have the same injection power in the LF). The only significant effect of increasing the post-Coriolis shock speed to $v_{\rm Cor} = 10^{10}$~cm~s$^{-1}$ is the increase of the adiabatic losses by a factor of $\sim 3$ at the Coriolis turnover location and beyond (not shown in the figures). This results in a decrease of $N'(E')$ for $E' \lesssim 100$~MeV, where adiabatic cooling (and particle escape) dominates.

\subsection{Spectral energy distribution}\label{SED}

\begin{figure}
    \centering
    \includegraphics[angle=270,width=\linewidth]{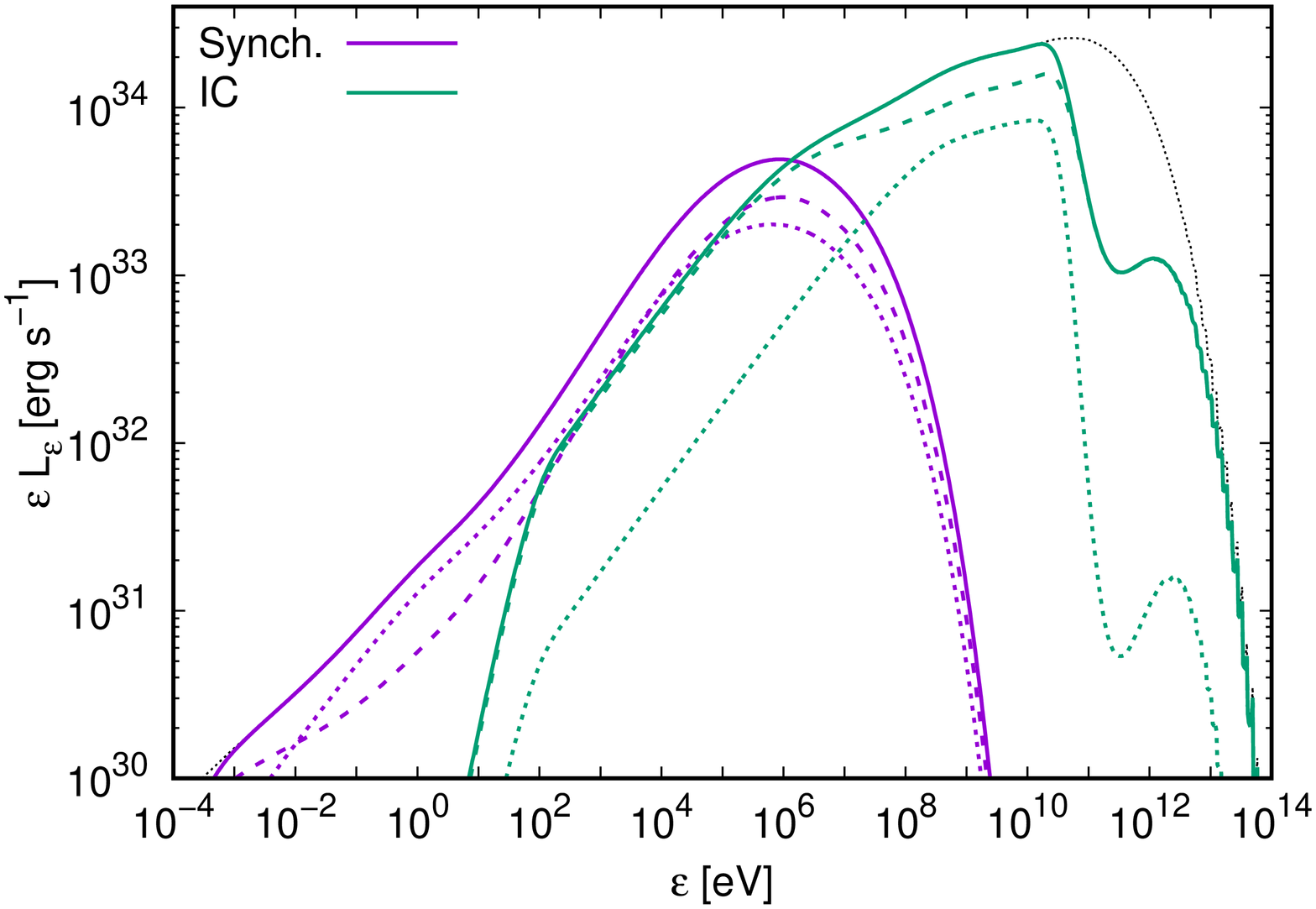}\\
    \vspace{3mm}
    \includegraphics[angle=270,width=\linewidth]{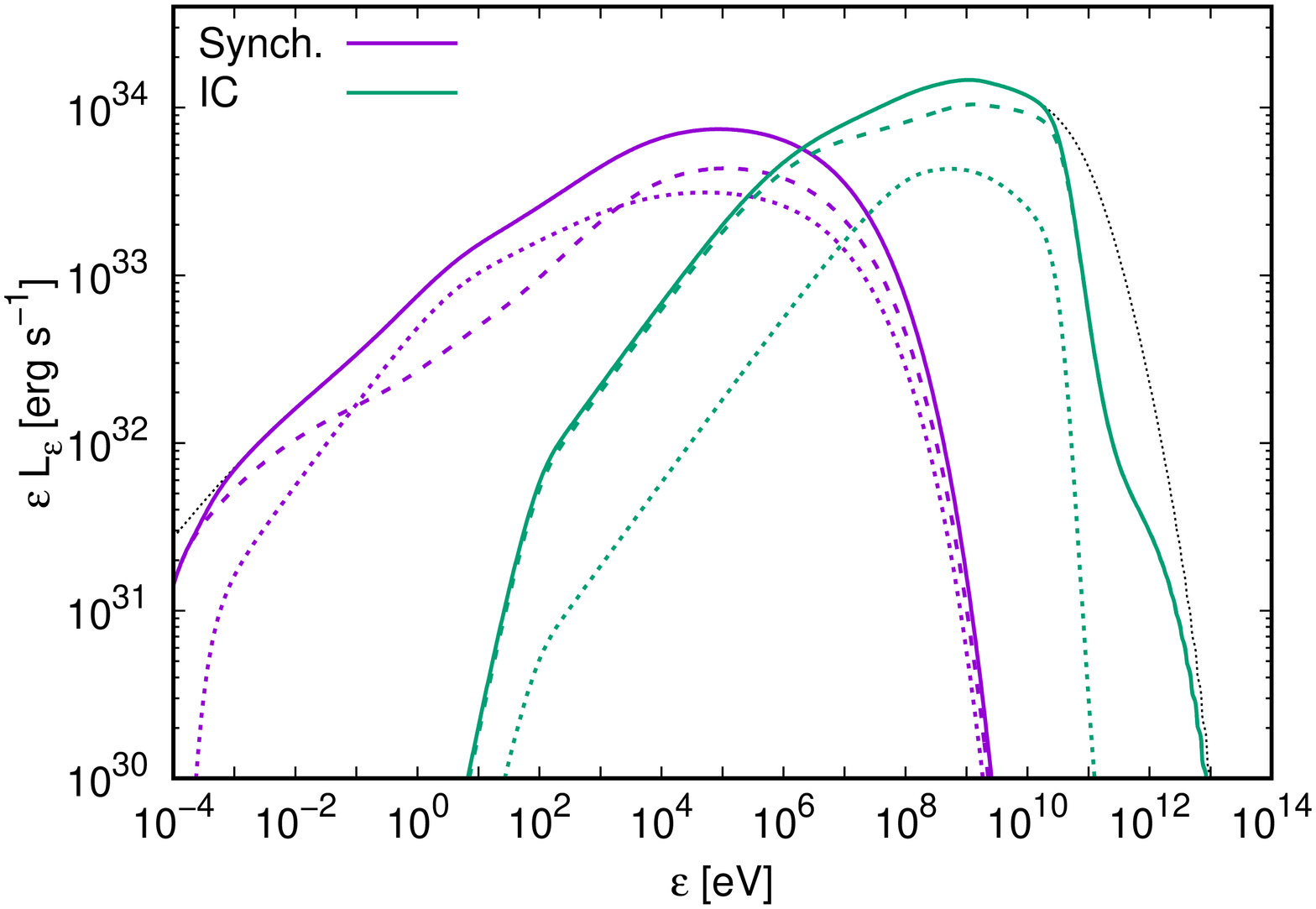}
    \caption{Observer synchrotron (purple lines) and IC (green lines) spectral energy distributions for $\Phi = 0.3$, $v_{\rm Cor} = 3\times10^9$~cm~s$^{-1}$, $i = 60\degree$, and $\eta_B = 10^{-3}$ (\textit{top panel}) and $10^{-1}$ (\textit{bottom panel}). The contributions of the inner and outer regions are represented with dotted and dashed lines, respectively. The black dotted lines show the total unabsorbed emission.}
    \label{fig:SEDv3}
\end{figure}

\begin{figure}
    \centering
    \includegraphics[angle=270,width=\linewidth]{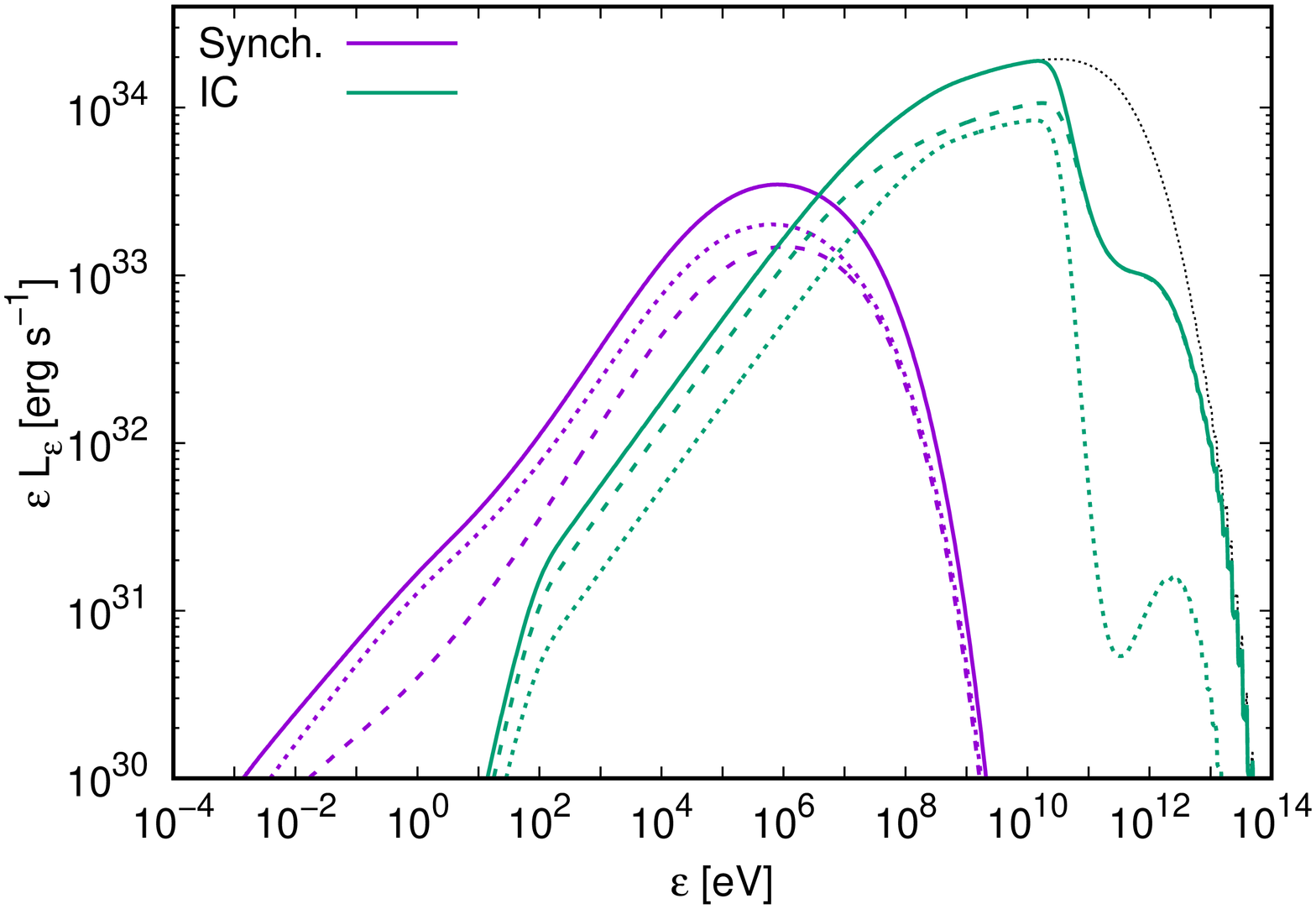}\\
    \vspace{3mm}
    \includegraphics[angle=270,width=\linewidth]{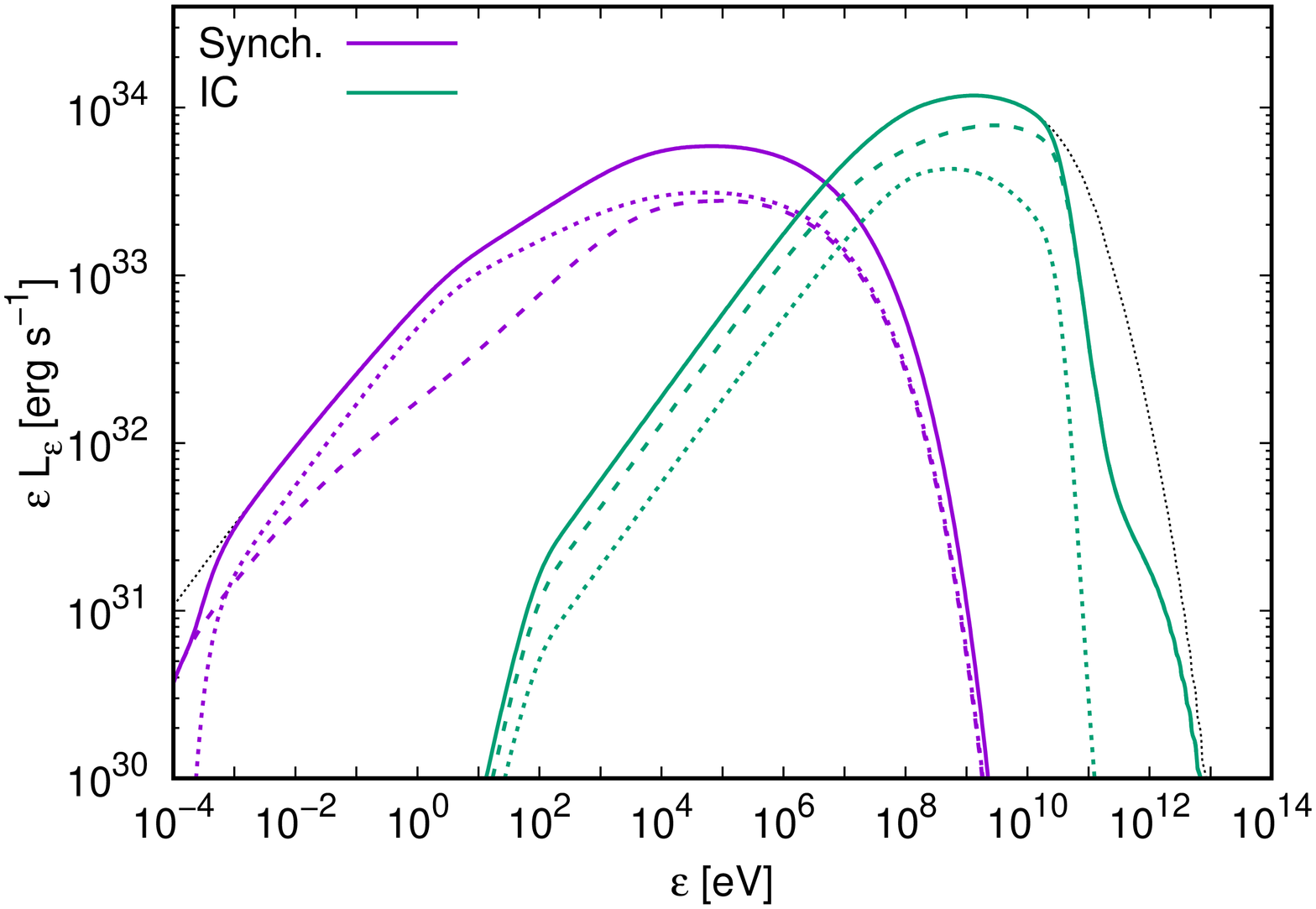}
    \caption{Same as in Fig.~\ref{fig:SEDv3}, but for $v_{\rm Cor} = 10^{10}$~cm~s$^{-1}$.}
    \label{fig:SEDv10}
\end{figure}

The synchrotron and IC SEDs, as seen by the observer, are shown in Figs.~\ref{fig:SEDv3} and \ref{fig:SEDv10} for $v_{\rm Cor} = 3\times10^9$~cm~s$^{-1}$ and $v_{\rm Cor} = 10^{10}$~cm~s$^{-1}$, respectively. We take a representative orbital phase of $\Phi = 0.3$, $i = 60\degree$, and $\eta_B = 10^{-3}$ and $10^{-1}$. The overall SED has the typical shape for synchrotron and IC emission, with the magnetic field changing the relative intensity of each component. The IC SED is totally dominated by the outer region even if it is farther from the star and the target photon field is less dense than in the inner region. This happens because the former contains many more (accumulated) non-thermal particles that scatter stellar photons (see Fig.~\ref{fig:NE}), and also because the ratio of synchrotron to IC cooling is smaller than in the inner region (therefore, more energy is emitted in the form of IC photons). Synchrotron radiation, on the other hand, is more equally distributed between the two regions. We note, however, that Doppler boosting could make the inner region dominate both the IC and synchrotron emission in a broad energy range for orbital phases close to the INFC (see Sect.~\ref{orbit}, and the discussion in Sect.~\ref{generalDiscussion}).

\subsection{Orbital variability}\label{orbit}

\begin{figure}
    \centering
    \includegraphics[angle=270,width=\linewidth]{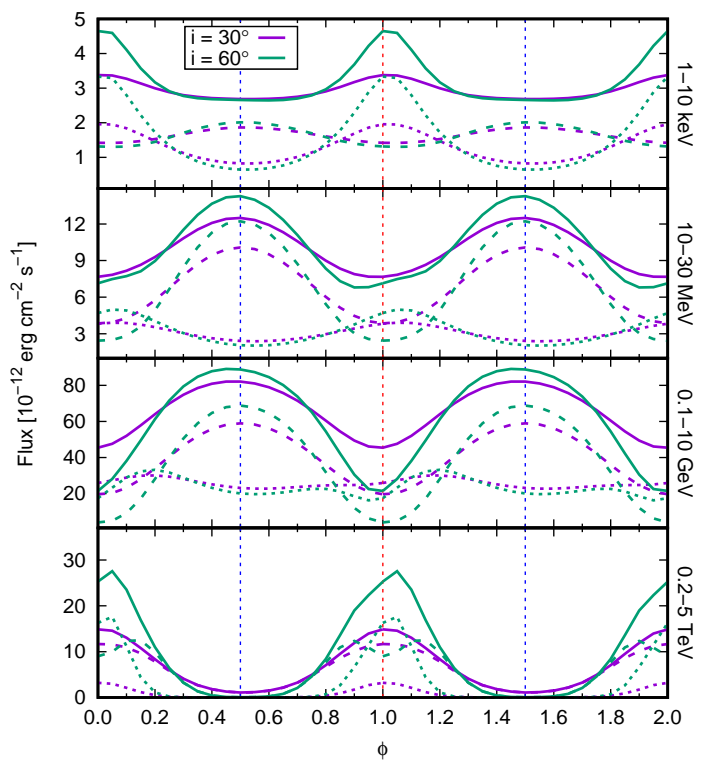}
    \caption{Light curves at different energy ranges (indicated in the right side) for $\eta_B = 10^{-3}$, $v_{\rm Cor} = 3\times10^9$~cm~s$^{-1}$, and $i = 30\degree$ (purple lines) and $60\degree$ (green lines). The contributions from the inner and outer regions are shown with dotted and dashed lines, respectively. The vertical dotted blue and red lines show the position of the superior and inferior conjunctions, respectively. Two orbits are represented for a better visualization.}
    \label{fig:LCv3_Blow}
\end{figure}

\begin{figure}
    \centering
    \includegraphics[angle=270,width=\linewidth]{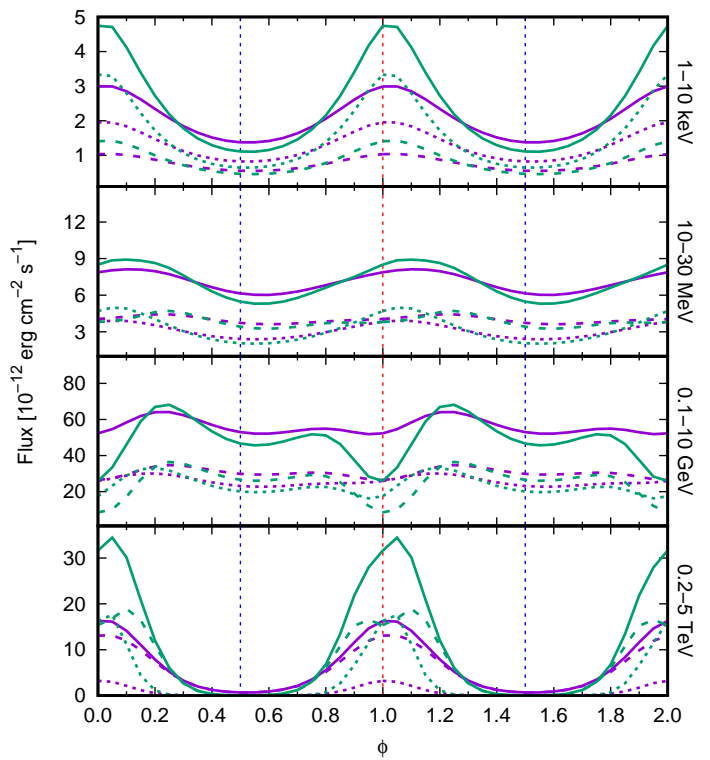}
    \caption{Same as in Fig.~\ref{fig:LCv3_Blow}, but for $v_{\rm Cor} = 10^{10}$~cm~s$^{-1}$.}
    \label{fig:LCv10_Blow}
\end{figure}

Light curves for two different system inclinations, $i = 30\degree$ and $60\degree$, and post-Coriolis shock speeds, $v_{\rm Cor} = 3\times 10^9$~cm~s$^{-1}$ and $10^{10}$~cm~s$^{-1}$, are shown in Figs.~\ref{fig:LCv3_Blow} and \ref{fig:LCv10_Blow} for $\eta_B = 10^{-3}$. Aside from a change in the flux normalization, the behavior of the light curves is very similar for $\eta_B = 10^{-1}$. The modulation of X-rays is correlated with that of VHE gamma rays (top and bottom panels, respectively). Low-energy (LE) gamma rays (second panel) show a correlated modulation with VHE gamma rays and X-rays for $v_{\rm Cor} = 10^{10}$~cm~s$^{-1}$, and an anti-correlated one for $v_{\rm Cor} = 3\times 10^9$~cm~s$^{-1}$. This change in the LE gamma-ray modulation is caused by a higher boosting (deboosting) of the outer region emission close to the INFC (SUPC) for $v_{\rm Cor} = 10^{10}$~cm~s$^{-1}$, which overcomes the intrinsic IC modulation. This same effect is responsible for high-energy (HE) gamma rays (third panel) to not show a clear correlation with other energy bands at high $v_{\rm Cor}$, whereas they are anti-correlated with VHE gamma rays and X-rays (and correlated with LE gamma rays) at low $v_{\rm Cor}$. The fact that Doppler boosting modulates the emission in the opposite way as IC does also causes the predicted variability in the inner region as seen by the observer to significantly decrease with respect to its intrinsic one, whereas the effect on the outer region is less extreme due to a lower fluid speed. Asymmetries can be observed in the light curves due to the spiral trajectory of the emitter, although they are mild because most of the radiation is emitted within a distance of a few orbital separations from the star, where the spiral pattern is just beginning to form. The asymmetry only becomes more noticeable for HE gamma rays, $i = 60\degree$, and $v_{\rm Cor} = 10^{10}$~cm~s$^{-1}$ (third panel in Fig.~\ref{fig:LCv10_Blow}), in which a double peak structure can be seen. The proximity to the star also makes VHE emission close to the SUPC to be almost suppressed by gamma-gamma absorption.

\begin{figure*}
    \centering
    \includegraphics[angle=270,width=\linewidth]{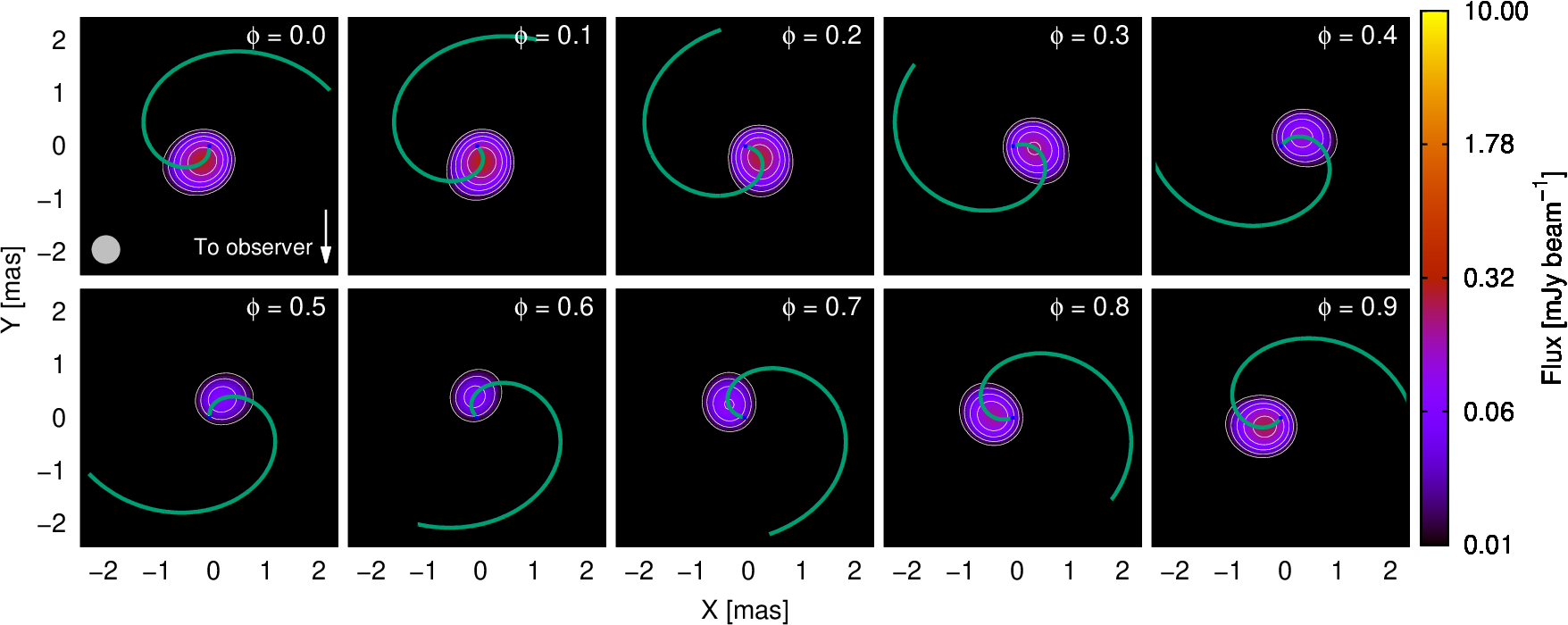}\\
    \vspace{3mm}
    \includegraphics[angle=270,width=\linewidth]{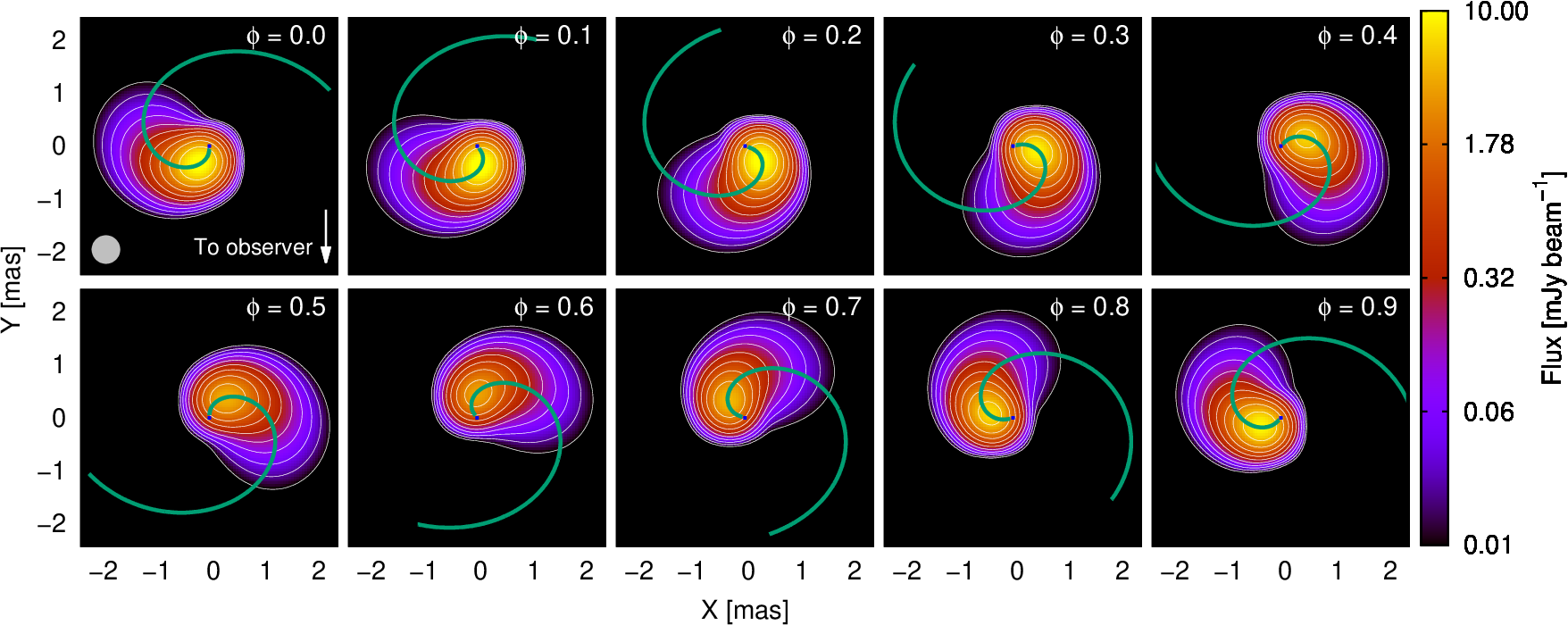}
    \caption{Simulated radio sky maps at 5~GHz for different orbital phases, $v_{\rm Cor} = 3\times10^9$~cm~s$^{-1}$, $i = 30\degree$, and $\eta_B = 10^{-3}$ (\textit{top panel}) and $10^{-1}$ (\textit{bottom panel}). The assumed telescope beam is shown as a gray circle in the bottom left corner of the first plot. The contour lines start at a flux of $10~\mu$Jy~beam$^{-1}$ and increase with a factor of 2. The star is represented (to scale) with a blue circle at (0,0), and the solid green line shows the axis of the conical emitter, the onset of which points towards the observer for $\Phi = 0$, and opposite to it for $\Phi = 0.5$.}
    \label{fig:skymap_i30}
\end{figure*}

\begin{figure*}
    \centering
    \includegraphics[angle=270,width=\linewidth]{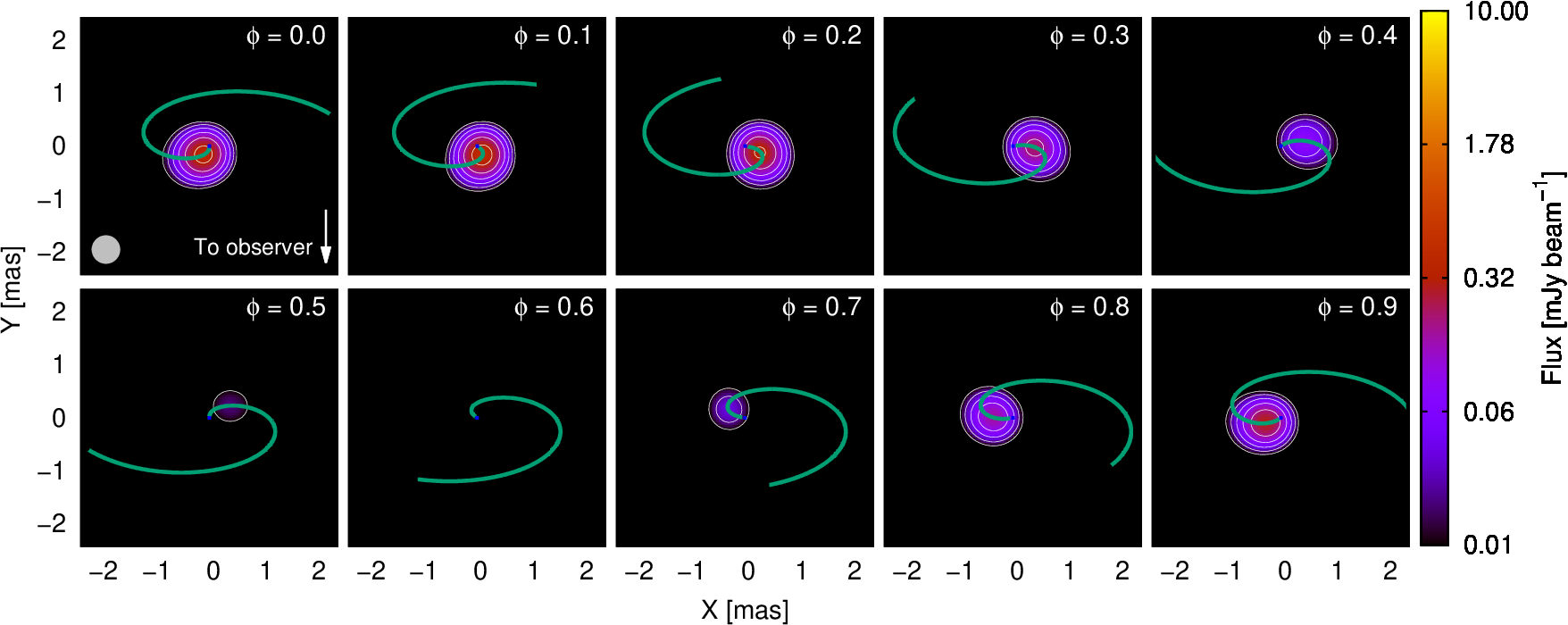}\\
    \vspace{3mm}
    \includegraphics[angle=270,width=\linewidth]{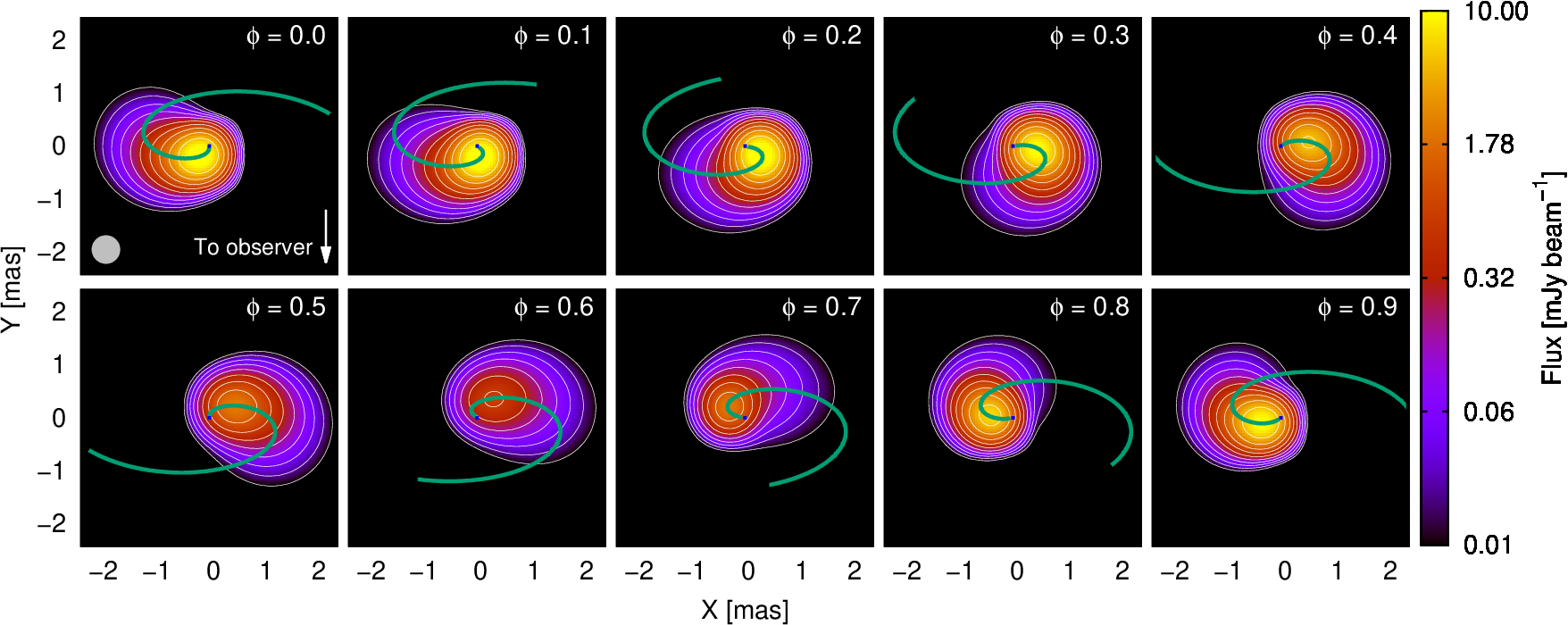}
    \caption{Same as in Fig.~\ref{fig:skymap_i30}, but for $i = 60\degree$.}
    \label{fig:skymap_i60}
\end{figure*}

Figures~\ref{fig:skymap_i30} and \ref{fig:skymap_i60} show, for $i = 30\degree$ and $60\degree$ respectively, simulated radio sky maps at 5~GHz for $v_{\rm Cor} = 3\times 10^9$~cm~s$^{-1}$, and $\eta_B = 10^{-3}$ and $10^{-1}$. They are obtained by convolving the projected emission of each segment with a 2D Gaussian with increasing width to approximately simulate the segment perpendicular extension. The resulting maps are then convolved again with a Gaussian telescope beam of FWHM = 0.5~mas. The contours are chosen so that the outermost one is $10~\mu$Jy~beam$^{-1}$, of the order of the sensitivity of very-long-baseline interferometry (VLBI). Due to free-free absorption, radio emission from the inner region is highly suppressed, and what can be seen in the sky maps comes mostly from the outer region. In particular, the part of the outer region that contributes most to the radio emission is located close to the Coriolis shock, and has an angular size of $\sim 0.5$~mas, which corresponds to a linear size of $\sim 1$~AU at the assumed distance of 3~kpc. For high $\eta_B$ and the assumed angular resolution, the initial part of the spiral structure can be traced in the radio images, especially for low inclinations. For small $\eta_B$, only the sites very close to the Coriolis shock contribute to the emission due to the low synchrotron efficiency farther away, and hints of a spiral outflow cannot be seen. Regardless of the magnetic field value, the position of the maximum of the radio emission shifts for as much as $\approx 1$~mas at different orbital phases.

\section{Application to LS 5039}\label{LS5039}

LS~5039 is a widely studied binary system hosting a main sequence O-type star and a compact companion, the nature of which is still unclear. Low system inclinations ($i \lesssim 40\degree$) favor a black hole scenario, whereas higher inclinations favor the compact object to be a neutron star. Inclinations above $\approx 60\degree$ are unlikely due to the absence of X-ray eclipses in this system \citep{casares05}. LS~5039 has an elliptical orbit with a semi-major axis of $a \approx 2.4\times10^{12}$~cm, an eccentricity of $e = 0.35 \pm 0.04$, and a period of $T \approx 3.9$~days. The superior and inferior conjunctions are located at $\Phi = 0.058$ and $0.716$, respectively, with $\Phi = 0$ corresponding to the periastron. The star has a luminosity of $L_\star = (7 \pm 1) \times10^{38}$~erg~s$^{-1}$, a radius of $R_\star = 9.3 \pm 0.7$~R$_\sun$, and an effective temperature of $T_\star = (3.9 \pm 0.2) \times 10^4$~K \citep{casares05}. The stellar mass-loss rate obtained through H$\alpha$ measurements is in the range $\dot{M}_{\rm w} = 3.7-4.8\times10^{-7}$~M$_\sun$~yr$^{-1}$ \citep{sarty11}, although this value would be overestimated if the wind were clumpy \citep[e.g.][]{muijres11}. The assumption of an extended X-ray emitter (as it is the case in this work) places an upper limit for $\dot{M}_{\rm w}$ of up to a few times $10^{-7}$~M$_\sun$~yr$^{-1}$, with the exact value depending on the system parameters (\citealt{szostek11}; see also \citealt{bosch07}). The lack of thermal X-rays in the shocked stellar wind, in the context of a semi-analytical model of the shocked wind structure, puts an upper limit in the putative pulsar spin down luminosity of $L_{\rm p} \le 6\times10^{36}$~erg~s$^{-1}$ \citep{zabalza11}. The latest \textit{Gaia} DR2 parallax data \citep{gaia18,luri18} sets a distance to the source of $d = 2.1 \pm 0.2$~kpc. The rest of the model parameters are unknown for LS~5039.

Carrying out a statistical analysis is hardly possible in our context, as we have many free parameters or parameters that are loosely constrained. Thus, we have looked for a set of parameter values that approximately reproduce the observational data, but the result should be considered just as illustrative of the model capability to reproduce the source behavior, and not a fit. We note that, for this purpose, we use a different value of the non-thermal power fraction for each accelerator ($\eta_{\rm NT}^A$ for the CD apex, and $\eta_{\rm NT}^B$ for the Coriolis shock). Table~\ref{tab:LS5039} shows all the model parameters used for the study of LS~5039, which are left constant throughout the whole orbit. For these parameters, we obtain $\eta = 0.035$, $\theta = 35.5\degree$, $0.55a \le r_{\rm apex} \le 1.14a$, and $1.03a \le r_{\rm Cor} \le 1.29a$, with the lower (upper) limits corresponding to the periastron (apastron).

\begin{table}
 \begin{center}
	\caption{Parameters used for the study of LS~5039.}
	\begin{tabular}{c l c}
    \hline \hline
	\multicolumn{1}{c}{}    & \multicolumn{1}{c}{Parameter} & Value                                 \\
    \hline
    \rule{0pt}{2.3ex}
	Star        & Temperature $T_\star$		   	            & $4\times10^4$~K                       \\
                & Luminosity $L_\star$ 		                & $7\times10^{38}$~erg~s$^{-1}$		    \\
                & Mass-loss rate $\dot{M}_{\rm w}$          & $1.5\times10^{-7}$~M$_\sun$~yr$^{-1}$	\\
                & Wind speed $v_{\rm w}$  	        	    & $3\times10^8$~cm~s$^{-1}$		        \\
    \hline
    \rule{0pt}{2.3ex}
    Pulsar      & Luminosity $L_{\rm p}$ 	            	& $3\times10^{36}$~erg~s$^{-1}$         \\
                & Wind Lorentz factor $\Gamma_{\rm p}$      & $10^5$                                \\
    \hline
    \rule{0pt}{2.3ex}
    System      & Orbit semi-major axis $a$ 		        & $2.4\times10^{12}$~cm                 \\
                & Orbital period $T$ 				        & $3.9$~days                            \\
                & Orbital eccentricity $e$                  & $0.35$                                \\
                & Distance to the observer $d$ 				& $2.1$~kpc                             \\
                & CD apex NT fraction $\eta_{\rm NT}^A$ 	& $0.03$                                \\
                & Cor. shock NT fraction $\eta_{\rm NT}^B$ 	& $0.18$                                 \\
                & Acceleration efficiency $\eta_{\rm acc}$  & $0.8$                                   \\
                & Injection power-law index $p$             & $-1.3$                                \\
                & Coriolis turnover speed $v_{\rm Cor}$     & $3\times10^9$~cm~s$^{-1}$             \\
                & Magnetic fraction $\eta_B$                & $0.02$                                \\
                & System inclination $i$                    & $40\degree$, $60\degree$              \\
    \hline
    \end{tabular}
    \label{tab:LS5039}
 \end{center}
\end{table}

\begin{figure}
    \centering
    \includegraphics[angle=270,width=\linewidth]{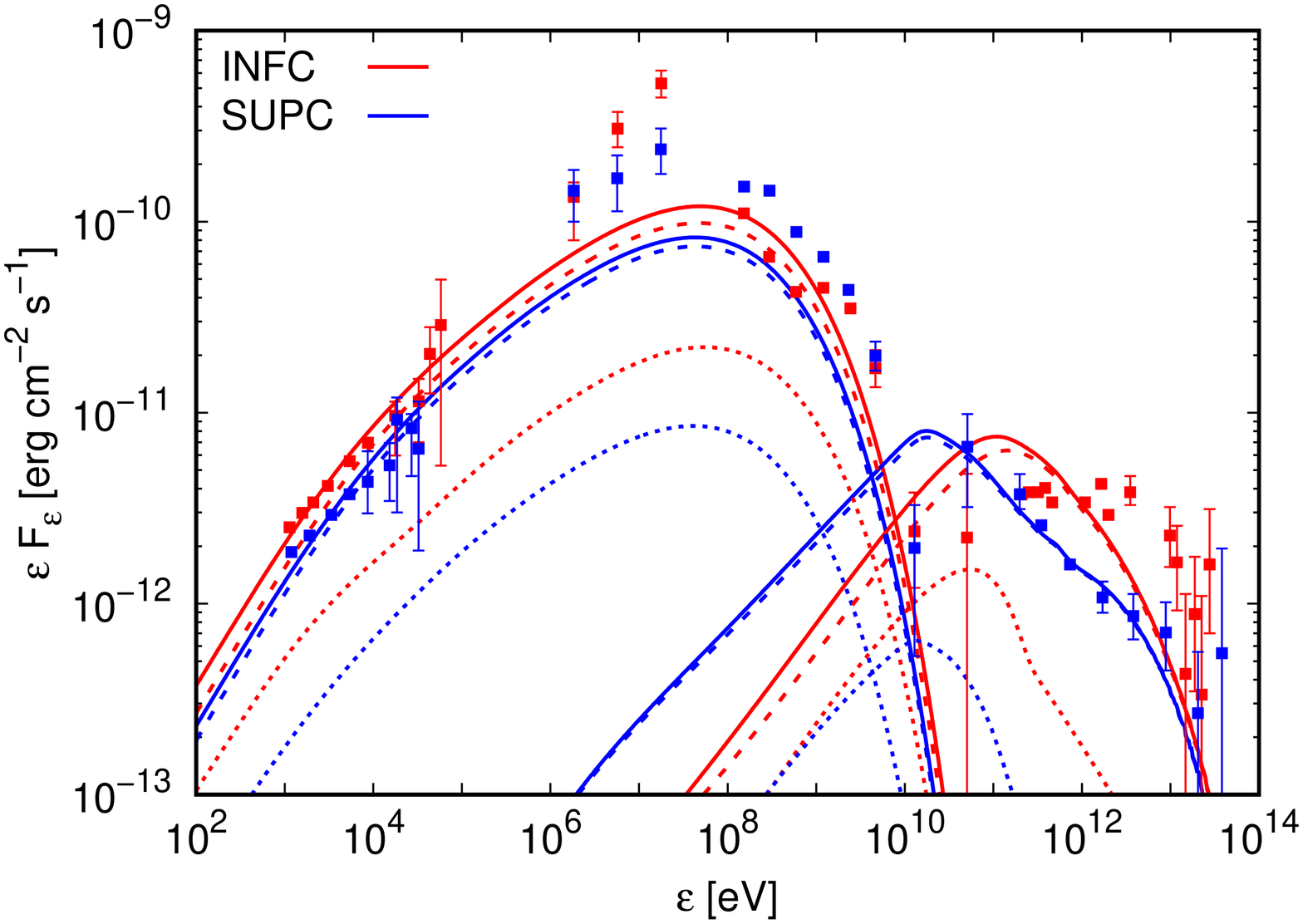}
    \caption{Observer synchrotron and IC SEDs of LS~5039 for $i = 60\degree$, averaged over the INFC (red lines; $0.45 < \Phi \le 0.90$) and SUPC (blue lines; $0.90 < \Phi$ or $\Phi \le 0.45$) phase intervals. Dotted and dashed lines represent the contributions of the inner and outer regions, respectively. From left to right, data from \textit{Suzaku}, COMPTEL, \textit{Fermi}/LAT, and H.E.S.S. are also represented.}
    \label{fig:LS5039_SED}
\end{figure}

Figure~\ref{fig:LS5039_SED} shows, for $i = 60\degree$, the computed SED averaged over two wide phase intervals, one around the INFC ($0.45 < \Phi \le 0.90$), and the other one around the SUPC ($0.90 < \Phi$ or $\Phi \le 0.45$). Observational data points of \textit{Suzaku} \citep{takahashi09}, COMPTEL \citep{collmar14}, \textit{Fermi}/LAT \citep{fermi09a,hadasch12}, and H.E.S.S. \citep{hess06} averaged over the same phase intervals are also plotted. The SEDs for different system inclinations in which a pulsar scenario is viable ($40\degree \lesssim i \lesssim 60\degree$) are not represented due their similarity to the one shown in Fig.\ref{fig:LS5039_SED}. Synchrotron emission dominates for $\varepsilon \lesssim 10$~GeV, with IC only contributing significantly at VHE. With the exception of the COMPTEL energies (1~MeV $\lesssim \varepsilon \lesssim$ 30~MeV), the SED reproduces reasonably well the magnitude of the observed fluxes, especially for X-rays and VHE gamma rays. At energies of 100~MeV $\lesssim \varepsilon \lesssim$ 10~GeV, the model overpredicts (underpredicts) the emission around INFC (SUPC). The hard electron spectrum allows the IC component not to strongly overestimate the fluxes around 10~GeV, although we must note that IC emission from secondary pairs, not taken into account, may increase a bit the predicted fluxes.

\begin{figure}
    \centering
    \includegraphics[angle=270,width=\linewidth]{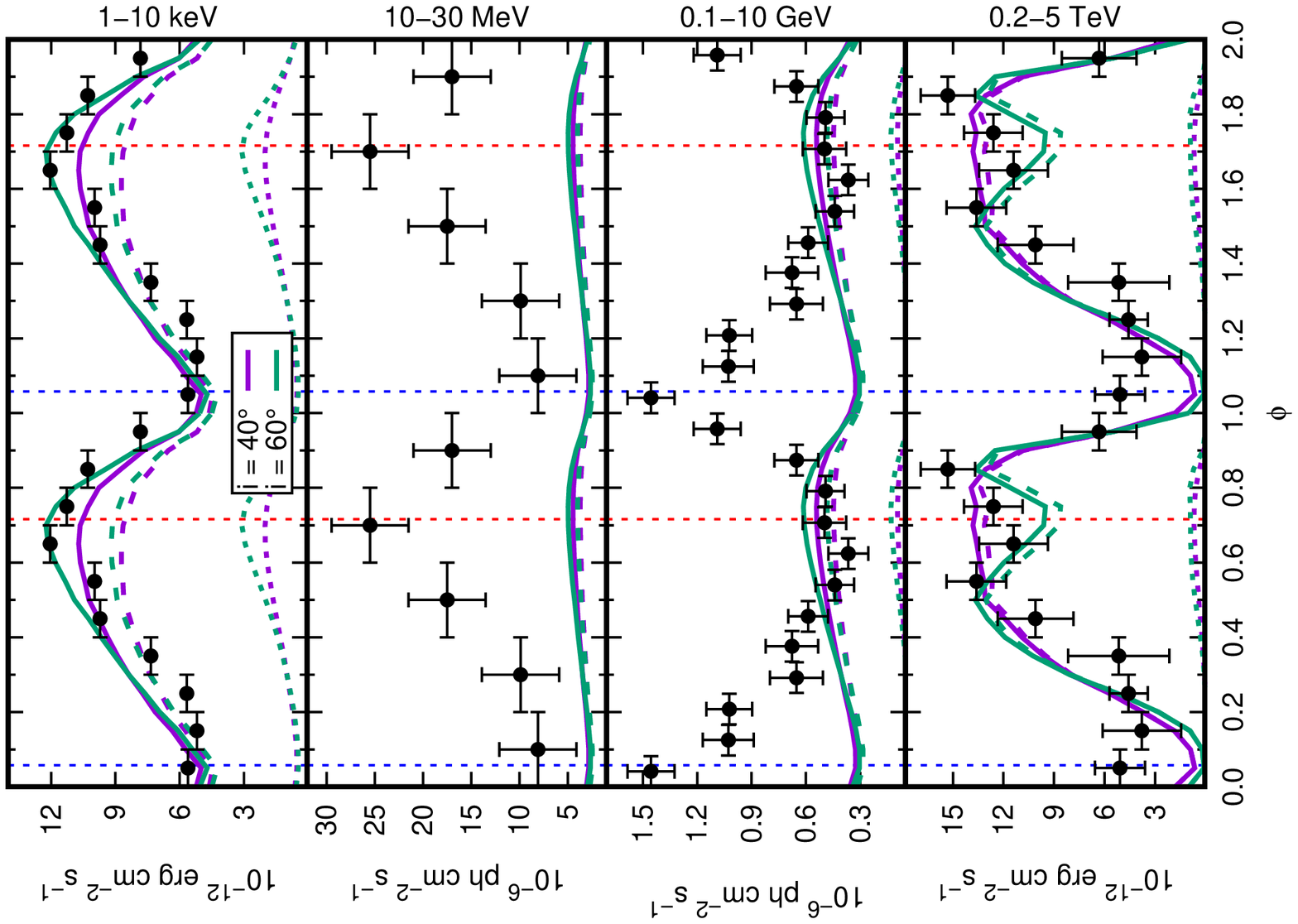}
    \caption{From \textit{top} to \textit{bottom}: Light curves of LS~5039 in the \textit{Suzaku} (1--10~keV), COMPTEL (10--30~MeV), \textit{Fermi}/LAT (0.1--10~GeV), and H.E.S.S. (0.2--5~TeV) energy ranges, for $i = 40\degree$ (purple lines) and $60\degree$ (green lines). The contributions of the inner and outer regions are shown with dotted and dashed lines, respectively. The phases corresponding to the INFC (SUPC) are shown with red (blue) vertical dashed lines. Note that the flux units are not the same for all the plots.}
    \label{fig:LS5039_LC}
\end{figure}

The computed LS~5039 light curves are shown in Fig.~\ref{fig:LS5039_LC} for the limits of the system inclination range allowed for a pulsar binary system ($40\degree \lesssim i \lesssim 60\degree$). Most of the emission comes from the outer region, regardless of the energy range. As in the SED, the model matches well the \textit{Suzaku} and H.E.S.S observations, except for an underestimation of the latter fluxes around the SUPC. Inclinations close to $60\degree$ are favored by the presence of a double peak in the VHE fluxes, which is not reproduced for lower values of $i$. This double peak is originated by the effect of the system orientation in the IC emission and Doppler boosting; the peak at $\Phi \approx 0.5$ has a higher intrinsic IC emission, but a lower boosting than the peak at $\Phi \approx 0.85$. At \textit{Fermi}/LAT energies, the model predicts a maximum in the light curve around the INFC and a minimum around the SUPC, contrary to what is observed. This happens because, in this energy range, the model emission is dominated by synchrotron radiation, which has a maximum around the INFC due to Doppler boosting. In the COMPTEL energy range, the relative behavior of the computed light curve is similar to the observations, although a factor of 4--5 lower.

\begin{figure*}
    \centering
    \includegraphics[angle=270,width=\linewidth]{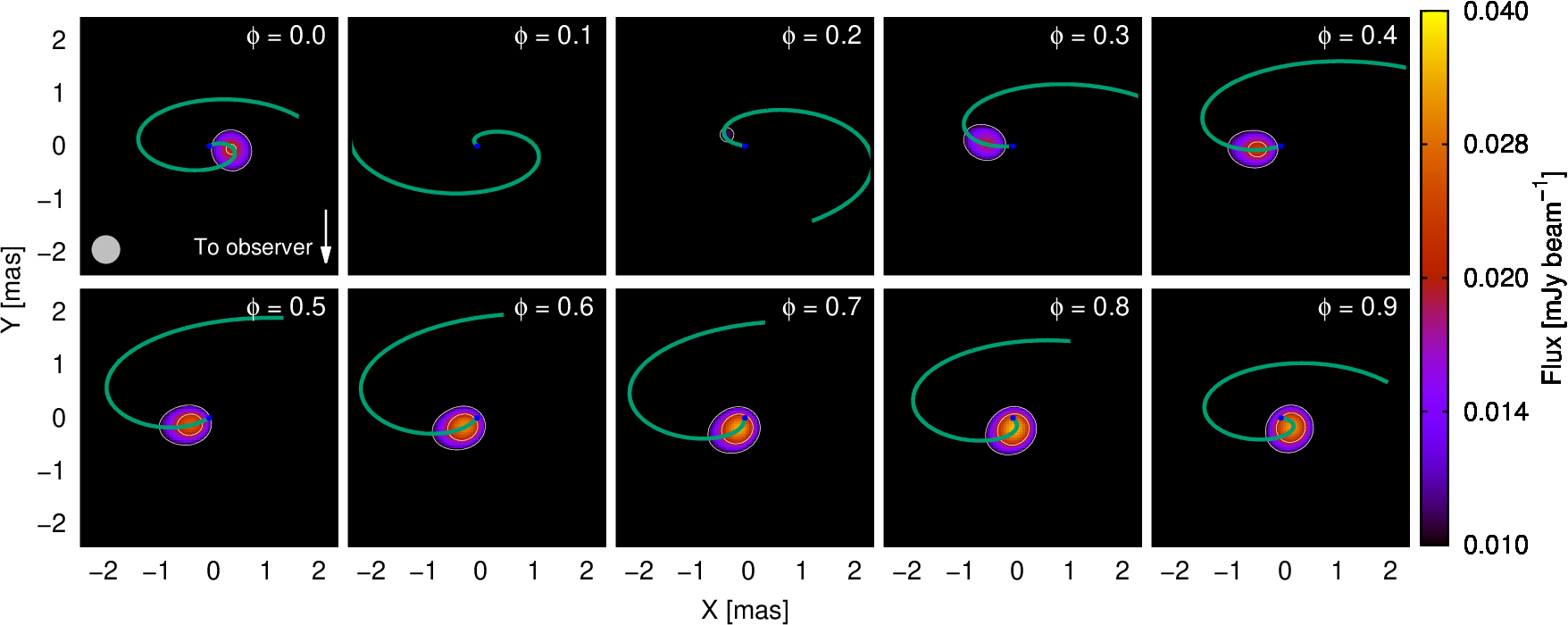}
    \caption{Same as in Fig.~\ref{fig:skymap_i60}, but for the LS~5039 model parameters, with $i = 60\degree$. Note the change in the color scale.}
    \label{fig:LS5039_skymap}
\end{figure*}

Figure~\ref{fig:LS5039_skymap} shows the computed radio sky map of LS~5039 at 5~GHz, for $i = 60\degree$ and a telescope beam with FWHM = 0.5~mas. The sky map for $i = 40\degree$ is very similar and is not shown. Since we do not consider particle reacceleration beyond the Coriolis shock, these maps show the synchrotron emission up to a few orbital separations from the latter, as farther away synchrotron emission is too weak to significantly contribute to the radio flux. This lack of reacceleration does not allow for a meaningful comparison between our model and the observations. While the predicted total radio flux at 5~GHz, averaged over a whole orbit, is 0.10~mJy, the detected one is around 20~mJy \citep{moldon12}. The assumption of a hard particle spectrum ($p = -1.3$) needed to explain the SEDs makes most of the available power to go into the most energetic electrons/positrons. Therefore, only a small part of the energy budget goes to those lower energy particles responsible for the radio emission, which makes the latter considerably fainter than in the generic case studied in Sect.~\ref{results}, in which $p = -2$. Despite the almost point-like and faint nature of the radio source, the emission maximum is displaced along the orbit by a similar angular distance of $\approx 1$~mas.

\section{Summary and discussion}\label{discussion}

We have developed a semi-analytical model, consisting of a 1D emitter, which can be used to describe both the dynamics and the radiation of gamma-ray binaries in a colliding wind scenario that includes orbital motion. In the following, we discuss the obtained results for a generic system and for the specific case of LS~5039, as well as the main sources of uncertainty.

\subsection{General case}\label{generalDiscussion}

In general, a favorable combination of non-radiative and radiative losses, and residence time of the emitting particles, leads to an outer emitting region that is more prominent in its non-thermal energy content and emission than the inner region. Thus, the SEDs are dominated by the outer region for most orbital phases. Moreover, both radio and VHE gamma-ray emission are suppressed close to the star due to free-free and gamma-gamma absorption, respectively, unless small system inclinations $i < 30\degree$ and/or orbital phases close to the INFC are considered. Nevertheless, the light curves show a non-negligible or even dominant contribution of the inner region close to the INFC for a broad energy range. This is mainly caused by the inner region emission being more Doppler-boosted than the outer region one, due to the flow moving faster in the former. Higher values of $v_{\rm Cor}$ (which could indicate a lower degree of mixing of stellar and pulsar winds) tend to decrease the radiative output of the outer region due to an increase in the adiabatic losses and in the escape rate of particles from the relevant emitting region. This trend is only broken for orbital phases close to the INFC, where a higher Doppler boosting is able to compensate for the decrease of intrinsic emission in the outer region. Doppler boosting (and hence the adopted velocity profile) has also a high influence on the orbital modulation of the IC radiation in both the inner and outer regions, to the point that emission peaks can become valleys owing to a change in the fluid speed of a factor of $\sim 3$ in the outer region\footnote{A similar effect is expected to happen if different velocity profiles are assumed for the inner region, but this has not been explicitly explored in this work}.

Radio emission could be used to track part of the spiral trajectory of the shocked flow for strong enough magnetic fields with $\eta_B \gtrsim 0.1$, while no evidence of such spiral structure is predicted for low fields. In any case, as long as the overall radio emission is detectable, variations of the image centroid position of the order of 1~mas (for a distance to the source of $\sim 3$~kpc) could be used as an indication of the dependency of the emitter structure with the orbital phase. This behavior is not exclusive of a colliding wind scenario, however, since the jets in a microquasar scenario could be affected by orbital motion in a similar manner \citep[e.g.][]{bosch13,molina18,molina19}.

There are some limitations in our model that should be acknowledged. One of them is the simplified dynamical treatment of the emitter, considered as a 1D structure. Not using proper hydrodynamical simulations makes the computed trajectory approximately valid within the first spiral turn, after which strong instabilities would significantly affect the fluid propagation, as seen in \cite{bosch15}. Nonetheless, since most of the emission comes from the regions close to the binary system, we do not expect strong variations in the radiative output due to this issue. We are not considering particle reacceleration beyond the Coriolis shock, although additional shocks and turbulence are expected to develop under the conditions present farther downstream \citep{bosch15}, and these processes could contribute to increase the non-thermal particle energetics. Accounting for reacceleration would in turn increase the emission farther from the binary system, resulting for instance in more extended radio structures, although with the aforementioned inaccuracies in the trajectory computation gaining more importance. 

Finally, we note that the model results for a generic case could change significantly for different values of some parameters that are difficult to determine accurately. The fluid speed and detailed geometry in the inner and outer regions are hard to constrain observationally. Precise values of the magnetic field, the electron injection index, the acceleration efficiency, and the non-thermal luminosity fraction are also difficult to obtain due to the presence of some degeneracy among them. Any significant changes in these quantities with respect to the values adopted in this work could have a strong influence on the emission outputs of the system.

\subsection{LS~5039}\label{LS5039Discussion}

Several model parameters can be fixed for the study of LS~5039 thanks to the existing observations of the source. Those parameters that cannot be obtained from observations are determined by heuristically (although quite thoroughly) exploring the parameter space, trying to better reproduce the observed LS~5039 emission. For this purpose, a hard particle spectrum is injected at both accelerators, with a power-law index $p = -1.3$, and a very high acceleration efficiency of $\eta_{\rm acc} = 0.8$ is assumed in both locations (we note that values between 0.5 and 1 give qualitatively similar results). These values are higher than those used for the general case, already quite extreme, although they cannot be discarded given the current lack of knowledge on the acceleration processes taking place in LS~5039 \citep[or even the emitting process; see e.g.][for an alternative origin of the VHE emission]{khangulyan20}. Additionally, since the outer region behavior reproduces better the observed (X-ray and VHE) light curves, a higher $\eta_{\rm NT}$ is assumed for the Coriolis shock than for the CD apex, providing the former with a larger non-thermal luminosity budget (i.e. $\eta_{\rm NT}=0.18$ versus 0.03).

The model nicely reproduces the observed X-ray and VHE gamma-ray emission of LS~5039, as well as the HE gamma-ray flux, although it fails to properly account for the HE gamma-ray modulation due to the synchrotron dominance in this energy range. A possible solution to this issue could be the inclusion of particle reacceleration beyond the Coriolis shock, which combined with the lower magnetic field may result in enough IC emission to explain the observed modulation of the HE gamma rays. This would also alleviate the extreme value of $\eta_{\rm acc}$ needed for the synchrotron emission to reach GeV energies. The largest difference between the model predictions and the observations comes at photon energies around 10~MeV, where the emission is underestimated by a factor of up to 5 (although interestingly the modulation is as observed). Some MeV flux would be added if we accounted for the synchrotron emission of electron-positron pairs created by gamma rays interacting with stellar photons \citep[e.g.][]{bosch08,cerutti10}, which is not included in our model. However, this component cannot be much larger than the TeV emission, as otherwise IC from these secondary particles would largely violate the 10~GeV observational flux constraints. Therefore, the energy budget of secondaries alone cannot explain the MeV flux.

The lack of predicted VHE emission around the SUPC is due to strong gamma-gamma absorption. The fact that we use a 1D emitter at the symmetry axis of the conical CD overestimates this absorption, since we are not considering emitting sites at the CD itself, which is approximately at a distance $R$ from the axis (see Fig.~\ref{fig:sketch}) or even farther (see the shocked pulsar wind extension in the direction normal to the orbital plane in the corresponding maps in \citealt{bosch15}). Within a few orbital separations from the pulsar (where most of the emission comes from), the CD is significantly farther from the star than its axis, and the observer, star, and flow relative positions are also quite different, allowing for lower absorption in some of the emitting regions. Therefore, the use of an extended emitter would reduce gamma-gamma absorption and increase the predicted VHE fluxes around the SUPC, possibly explaining the H.E.S.S. fluxes. Particle reacceleration beyond the Coriolis shock may also make some regions farther away from the star to emit VHE photons that would be less absorbed. As already discussed for the general case, reacceleration would also extend the radio emission and, if accounted for, could allow for a sky map comparison with VLBI observations \citep[e.g.][]{moldon12}. Thus, one could constrain a reacceleration region added to our model using VLBI observations, but this is beyond the scope of the present work.

Our computed SED is qualitatively different to the best fit scenario presented in \cite{delpalacio15}, in which a one-zone model was applied for an accelerator in a fixed position at a distance of $1.4a \approx 3.3\times10^{12}$~cm from the star. Although the general shape is similar, their SED is totally dominated by IC down to $\sim 1$~MeV, whereas in our model IC is only relevant at energies $\gtrsim 10$~GeV. Another one-zone model in \cite{takahashi09} explains well the X-ray and VHE gamma-ray emission and modulation, although it underestimates the fluxes at MeV and GeV energies (which were not available by the time of publication of this work). Our synchrotron and IC phase-averaged SEDs above $100$~MeV are somewhat close to those in \cite{zabalza13}. They applied a two-zone model to LS~5039, with the two emitting regions located at the CD apex and the Coriolis shock. Their model reproduces better the GeV modulation, although it also fails to explain the MeV emission, and underestimates the X-ray emission by more than one order of magnitude. Similar results were obtained by \cite{dubus15} with a model that computes the flow evolution through a 3D hydrodynamical simulation of the shocked wind close to the binary system, where orbital motion is still unimportant (and thus, it does not include the Coriolis shock). Modulation in the HE band is well explained there, although both the X-ray and the MeV emission are underestimated. All of the above, added to the fact that the 1D emitter model presented in this work also fails to reproduce some of the LS~5039 features, seems to point towards the need of more complex models to describe the behavior of this source, accounting for particle reacceleration, using data from 3D (magneto-)hydrodynamical simulations to compute the evolution of an emitter affected by orbital motion, and possibly including the unshocked pulsar wind zone to correctly describe the MeV radiation \citep[see, e.g.,][]{derishev12}.

\begin{acknowledgements}
We would like to thank the referee for his/her constructive and useful comments, which were helpful to improve the manuscript. We acknowledge support by the Spanish Ministerio de Econom\'{i}a y Competitividad (MINECO/FEDER, UE) under grant AYA2016-76012-C3-1-P, with partial support by the European Regional Development Fund (ERDF/FEDER), and from the Catalan DEC grant 2017 SGR 643. EM acknowledges support from MINECO through grant BES-2016-076342. 
\end{acknowledgements}

\FloatBarrier
\bibliographystyle{aa}
\bibliography{references}

\end{document}